\documentclass{cimento}
\pdfoutput=1

\usepackage{graphicx}

\title{The IceCube low-energy excess: a Dark Matter interpretation}
\author{Marco~Chianese}
\instlist{\inst{}Dipartimento di Fisica, Universit\`a di Napoli ``Federico II", Complesso Univ. Monte S. Angelo, Via Cinthia, I-80126 Napoli, Italy.  \inst{}  INFN, Sezione di Napoli, Complesso Univ. Monte S. Angelo, Via Cinthia, I-80126 Napoli, Italy.}

\begin{document}

\maketitle

\begin{abstract}
The recent study on the the 6-year up-going muon neutrinos by the IceCube Collaboration support the hypothesis of a two-component scenario explaining the diffuse TeV-PeV neutrino flux. Once a hard astrophysical power-law is considered, an excess in the IceCube data is shown in the energy range 10--100 TeV ({\it low-energy excess}). By means of a statistical analysis on the neutrino energy spectrum and on the angular distribution of neutrino arrival directions, we characterize a two-component neutrino flux where decaying/annihilating Dark Matter particles provide a contribution to the IceCube observations.
\end{abstract}

\section{Introduction}

The IceCube Neutrino Telescope has observed for the first time a diffuse extraterrestrial neutrino flux in the TeV--PeV energy range, with a deviation from the atmospheric background of about $7\sigma$~\cite{Aartsen:2014muf,Aartsen:2015knd,Aartsen:2015zva,Aartsen:2016xlq}. However, until now the origin of such a diffuse neutrino flux is unclear. Under the reasonable assumption of a correlation with hadronic cosmic-rays, one would expect that standard astrophysical sources give rise to a neutrino flux parametrized in terms of a power-law behavior $E_\nu^{-\gamma}$ with $\gamma$ being the {\it spectral index}~\cite{Loeb:2006tw,Kelner:2006tc,Winter:2013cla}.

The recent IceCube observations of 6-year up-going muon neutrinos~\cite{Aartsen:2016xlq} are well explained at high energies ($E_\nu \geq 100$~TeV) by a single hard power-law with $\gamma = 2.13\pm0.13$. Such a value is in a $3.3\sigma$ tension with the previous analyses that provide a combined best-fit spectral index of $2.50 \pm 0.09$~\cite{Aartsen:2015knd}. This tension suggests the presence of a second component in the 10--100~TeV energy range. Moreover, such a new component may have a galactic origin since the 6-year up-going muon neutrino data do not point towards the Galactic Center of the Milky Way.

Indeed, once an astrophysical power-law flux with spectral index 2.0 (2.2) is considered for all neutrino flavours\footnote{In case of standard astrophysical sources, the flavour ratio at the Earth is $\left(\nu_e:\nu_\mu:\nu_\tau\right) = \left(1:1:1\right)$ due to the neutrino oscillations.}, a {\it low-energy excess} (10--100~TeV) appears in both 2-year MESE~\cite{Aartsen:2014muf} and 4-year HESE~\cite{Aartsen:2015zva} IceCube data with a local statistical significance of $2.3\sigma$ ($1.9\sigma$)~\cite{Chianese:2016opp,Chianese:2016kpu}. Assuming that the low-energy excess is not just a statistical fluctuation, we have characterized the properties of a Dark Matter (DM) signal able to explain it. 

\section{Two-component neutrino flux}

In addition to the atmospheric background, we have proposed the following two-component neutrino flux
\begin{equation}
\frac{{\rm d}\phi}{{\rm d}E_\nu {\rm d}\Omega} = \phi_0^{\rm Astro}\left(\frac{E_\nu}{100\,\rm TeV}\right)^{-\gamma} + \frac{{\rm d}\phi^{\rm DM}}{{\rm d}E_\nu {\rm d}\Omega}\,,\label{eq:2c}
\end{equation}
where the DM neutrino flux takes the form\footnote{More details about the expressions of the DM flux can be found in ref.~\cite{Chianese:2016kpu}.}
\begin{eqnarray}
\left.\frac{{\rm d}\phi^{\rm DM}}{{\rm d}E_\nu {\rm d}\Omega}\right|_{\rm dec.} &=& \frac{1}{4\pi \, m_{\rm DM} \, \tau_{\rm DM}} \left\{f^{\rm G}_{\rm dec.} \left[\rho_h\left(s,\theta\right),\frac{{\rm d}N}{{\rm d}E_\nu}\right] + f^{\rm EG}_{\rm dec.} \left[\frac{{\rm d}N}{{\rm d}E_\nu}\right]\right\} \,,\label{eq:dec}\\
\left.\frac{{\rm d}\phi^{\rm DM}}{{\rm d}E_\nu {\rm d}\Omega}\right|_{\rm ann.} &=&\frac{\left< \sigma v \right>}{8\pi \, m^2_{\rm DM}} \left\{f^{\rm G}_{\rm ann.} \left[\rho_h\left(s,\theta\right)^2,\frac{{\rm d}N}{{\rm d}E_\nu}\right] + f^{\rm EG}_{\rm ann.} \left[\frac{{\rm d}N}{{\rm d}E_\nu},B\left(z\right)\right]\right\}\,,\label{eq:ann}
\end{eqnarray}
in the decaying (dec.) and annihilating (ann.) cases, respectively. In the above expressions, $m_{\rm DM}$ is the DM mass, whereas $\tau_{\rm DM}$ and $\left< \sigma v \right>$ are the lifetime and the thermally averaged cross-section, respectively. In the brackets, the Galactic component $f^{\rm G}$ depends on the angular coordinate $\theta$ measuring the angular distance from the  Galactic Center through the DM halo density profile $\rho_h$~\cite{Cirelli:2010xx}\footnote{We consider the Navarro-Frenk-White (NFW) distribution~\cite{Navarro:1995iw} as a benchmark.}, while the ExtraGalactic one $f^{\rm EG}$ is isotropic. The behaviour of the DM neutrino flux as a function of the energy is instead described by the energy spectrum ${\rm d}N/{\rm d}E_\nu$ that depends on the particular decaying/annihilating channel considered. Finally, in case of annihilating DM the ExtraGalactic component also depends on the so-called boost factor $B\left(z\right)$~\cite{Cirelli:2010xx}.

In order to infer the properties of the DM neutrino flux explaining the low-energy excess, we have performed two complementary studies: an {\it angular} analysis and an {\it energetic} one. The angular analysis is based on comparing the distribution of the arrival directions of IceCube events with the angular distributions expected from a DM signal. Since decaying and annihilating DM fluxes have distinct angular distributions due to the different dependence on the DM halo density profile (see eq.s~(\ref{eq:dec})~and~(\ref{eq:ann})), such an angular analysis can discern among the two DM signals. On the other hand, the analysis on the neutrino energy spectrum is sensitive to the decaying/annihilating channel considered since, for instance, the energy spectrum is quite different in case of leptons or quarks in the final-states. Moreover, such an analysis also provides the allowed regions in the parameters spaces $m_{\rm DM}$--$\tau_{\rm DM}$ and $m_{\rm DM}$--$\left< \sigma v \right>$ compatible with the IceCube observations.

\section{Results and conclusions}

\begin{figure}[t!]
\centering
\includegraphics[width=0.42\textwidth]{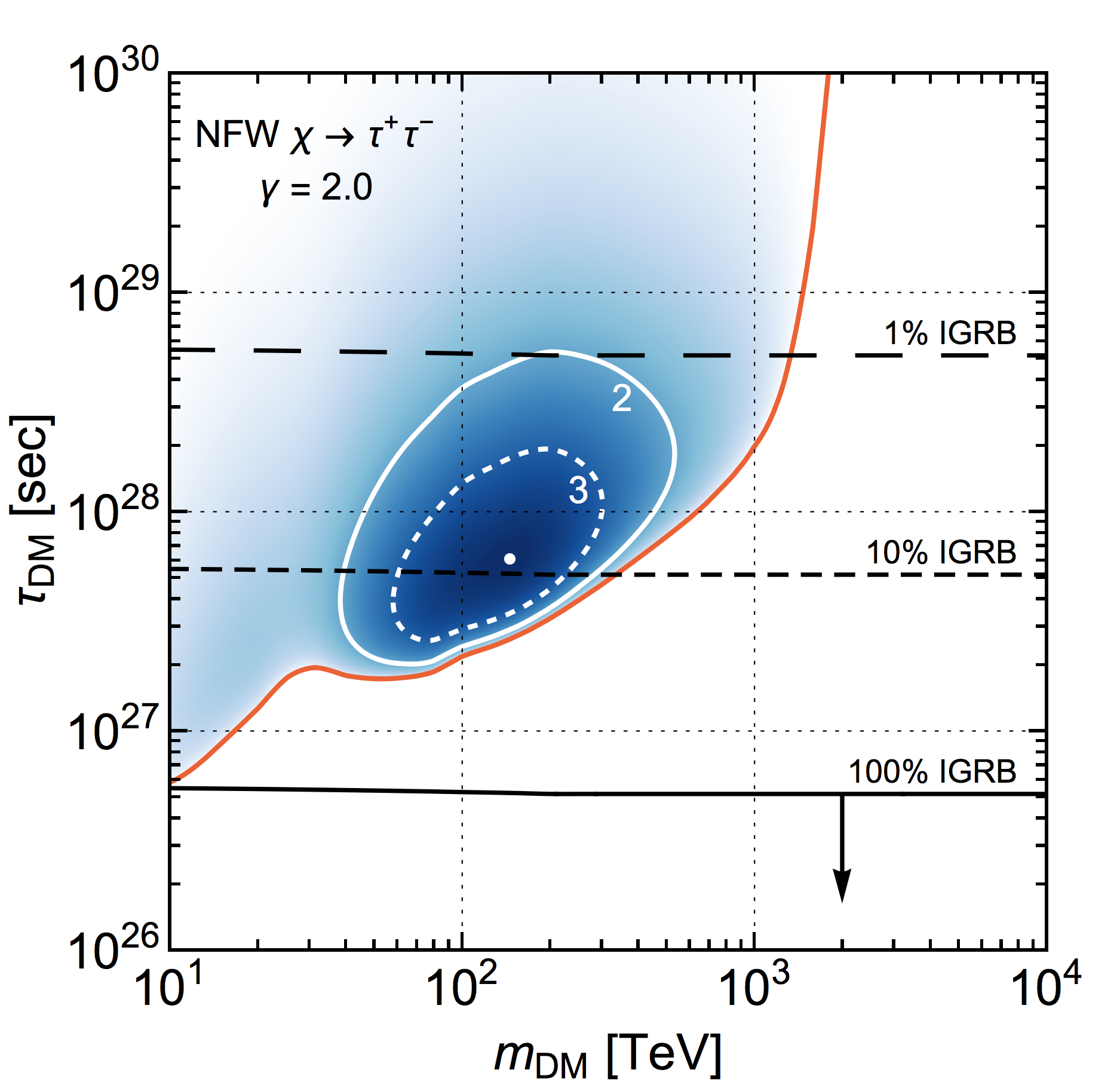}
\hskip3.mm
\includegraphics[width=0.42\textwidth]{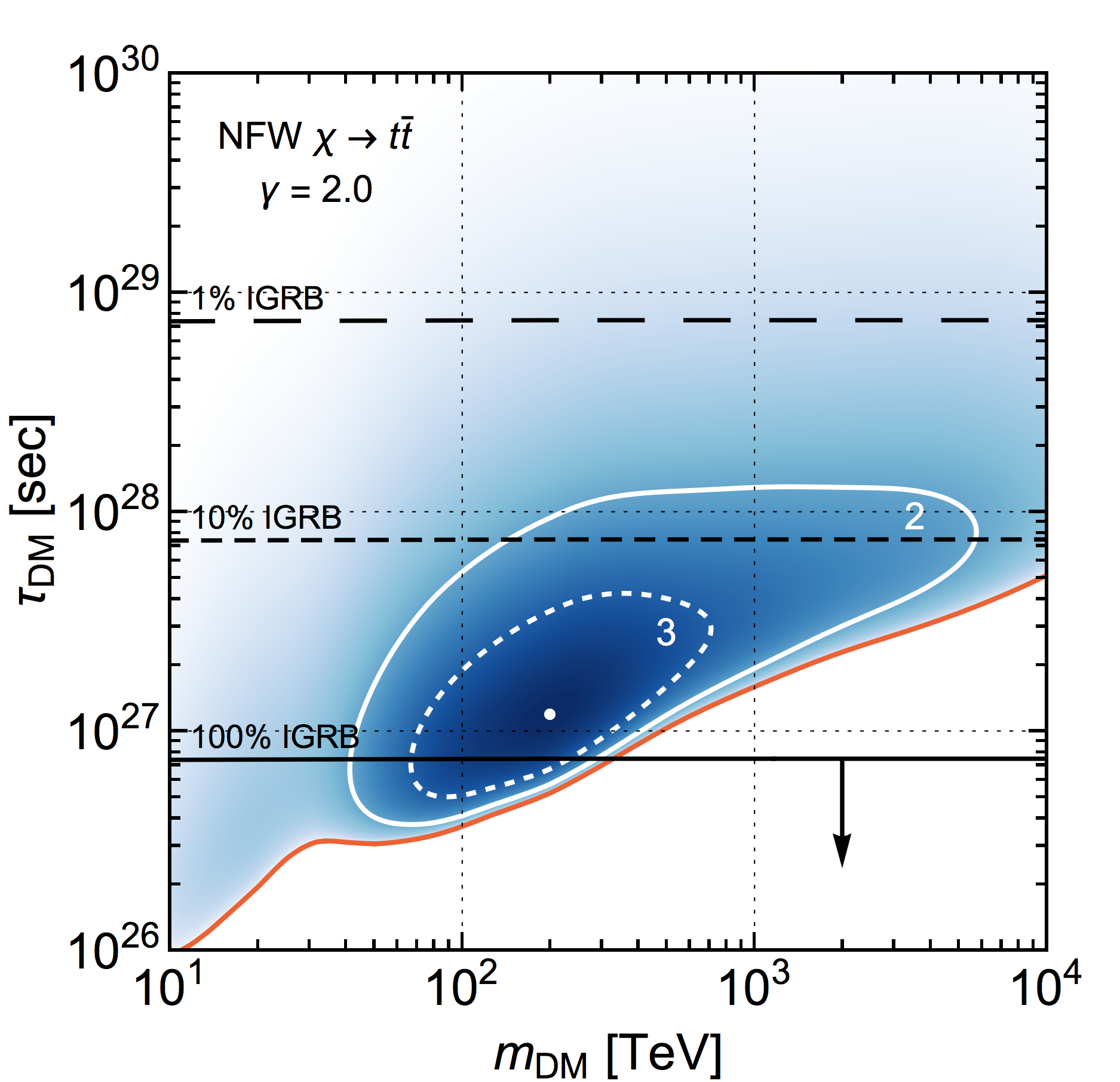}
\includegraphics[width=0.077\textwidth]{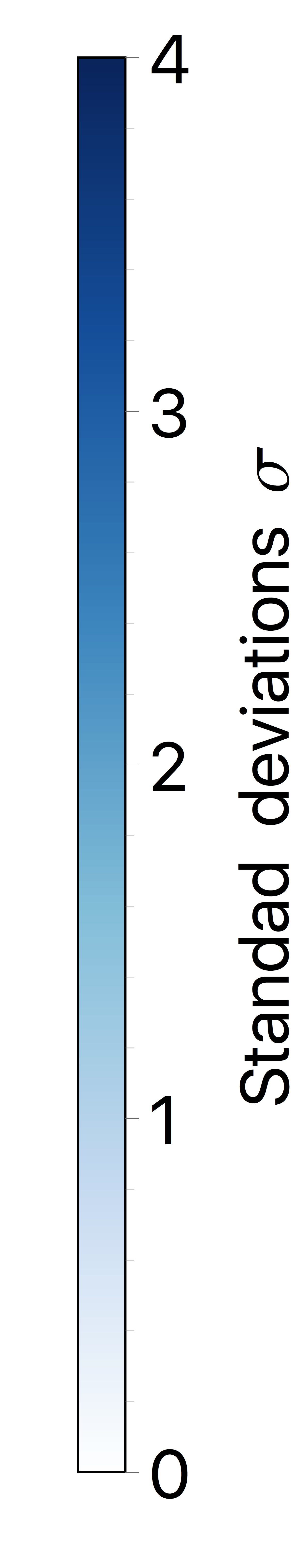}
\includegraphics[width=0.42\textwidth]{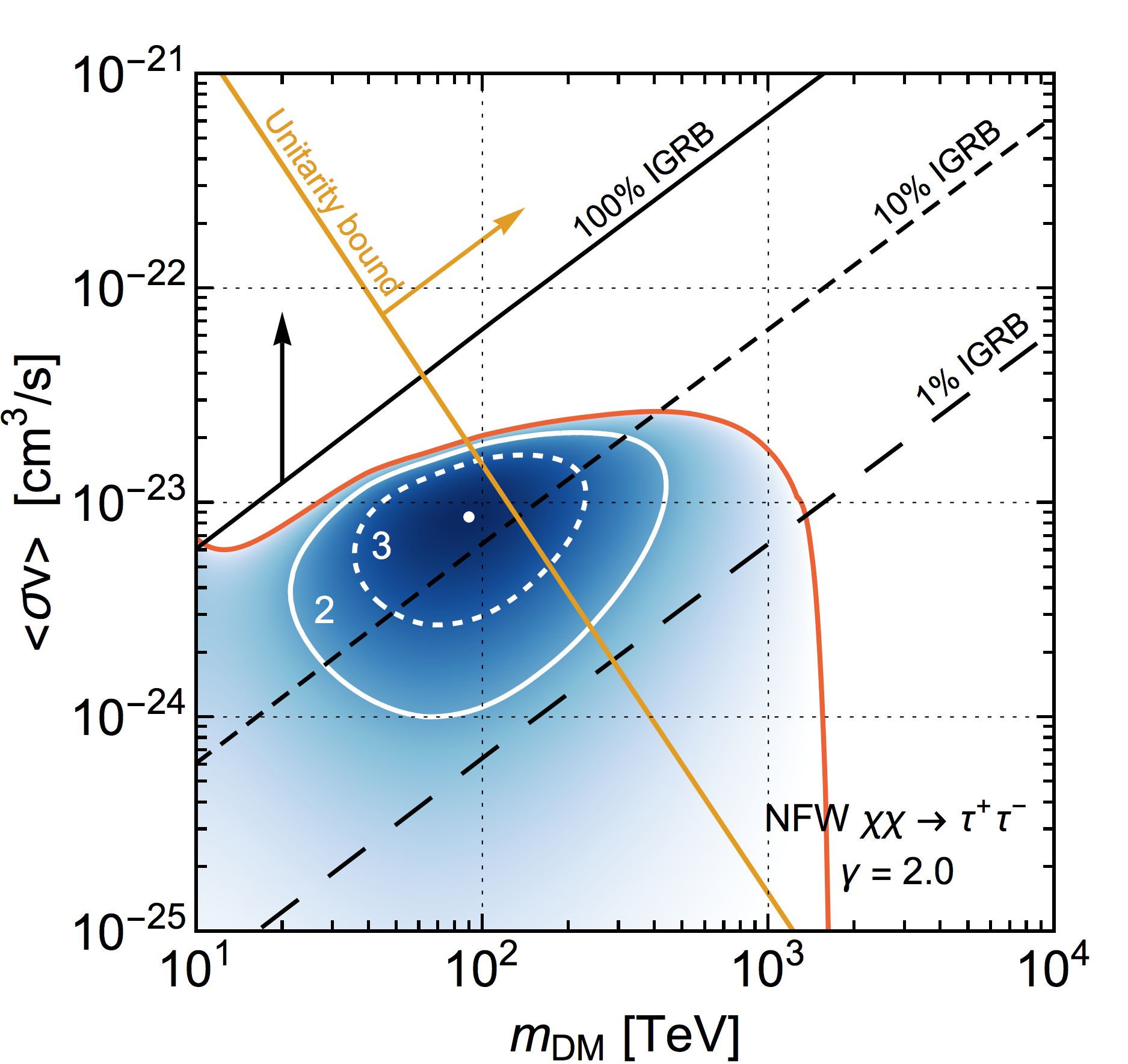}
\hskip3.mm
\includegraphics[width=0.42\textwidth]{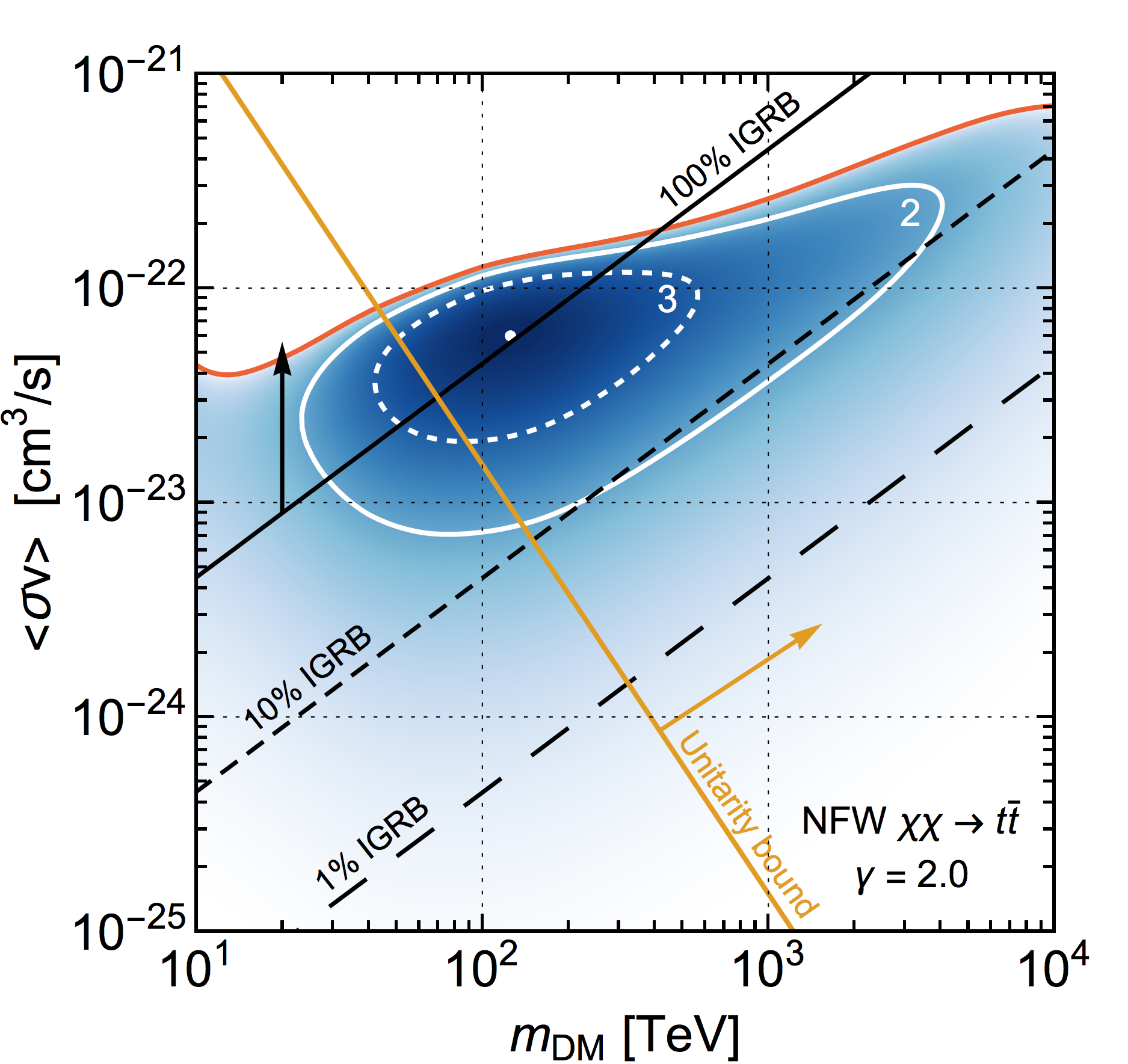}
\includegraphics[width=0.077\textwidth]{bar.png}
\caption{\label{fig:1}Statistical significance of the two-component scenario with respect the case of a single astrophysical power-law, once a spectral index of 2.0 is considered. The left (right) panels refer to the case of leptonic (hadronic) final-states, whereas the top (bottom) plots correspond to the decaying (annihilating) DM scenario. The plots are taken from ref.~\cite{Chianese:2016kpu}.}
\end{figure}
The angular analysis performed on the 4-year HESE data~\cite{Chianese:2016opp} shows that only the case of annihilating DM with a small boost factor is already ruled out, while other DM scenario are still allowed by data. Indeed, in order to statistically rule out a DM interpretation of the excess, hundreds of events in the 10--100~TeV energy range are required~\cite{Chianese:2016opp}.

The main results of the analysis on the neutrino energy spectrum are reported in fig.~\ref{fig:1} (see ref.~\cite{Chianese:2016kpu} for more details). The plots show the statistical preference in standard deviations $\sigma$ (evaluated by means of a Likelihood-ratio statistical test) of the IceCube data for the two-components scenario provided in eq.~(\ref{eq:2c}). The maximum significance of the DM component (white dot) reaches about $4\sigma$. Moreover, the red lines delimit from above the region in the parameter space that is excluded by the IceCube measurements. The gamma-rays constraints are shown by the black lines, which are related to different DM contributions (1\%, 10\% and 100\%) to the Isotropic diffuse Gamma-Ray Background (IGRB) measured by Fermi-LAT~\cite{Ackermann:2014usa}. Finally, the yellow lines bound the region that is not allowed according to the unitarity constraint on the cross-section~\cite{Griest:1989wd}.

It is worth observing that, since it is reasonable to assume that standard astrophysical sources account at least for the 90\% of the IGRB spectrum, the neutrino and gamma-ray data favour a decaying DM interpretation of the IceCube low-energy excess over a annihilating one. Furthermore, leptonic final-states (represented by $\tau^+\tau^-$ channel) are preferred with respect to hadronic ones (represented by $t\overline{t}$ channel).

\section*{\bf Acknowledgments} 

We acknowledge support by the Istituto Nazionale di Fisica Nucleare I.S. TASP and the PRIN 2012 “Theoretical Astroparticle Physics” of the Italian Ministero dell’Istruzione, Universit\`a e Ricerca.

\end{document}